\documentclass[a4paper]{article}

\usepackage{graphicx}
\usepackage{natbib}

\usepackage{my_definitions}

\usepackage{amssymb}

\title{The dependence of relative dispersion on turbulence scales in
Lagrangian Stochastic Models}
\author{A. Maurizi$^1$, G. Pagnini$^{1,2}$ and F. Tampieri$^1$\\
\small{$^1$ISAC-CNR, via Gobetti 101, I-40129 Bologna, Italy}\\
\small{$^2$Facolt\`a di Scienze Ambientali, Universit\`a di Urbino,}\\
\small{Campus Scientifico Sogesta, I-61029 Urbino, Italy}}

\begin{document}
\maketitle

\begin{abstract}
The aim of the article is to investigate the relative dispersion
properties of the Well Mixed class of Lagrangian Stochastic Models.
Dimensional analysis shows that given a
model in the class, its properties depend solely on a non-dimensional
parameter, which measures the relative weight of Lagrangian-to-Eulerian
scales. This parameter is formulated in terms of Kolmogorov constants,
and model properties are then studied by modifying its value in a range that
contains the experimental variability. Large variations are found
for the quantity $g^*=2gC_0^{-1}$, where $g$ is the Richardson constant,
and for the duration of the $t^3$ regime. Asymptotic analysis of model
behaviour clarifies some inconsistencies in the literature and excludes the
Ornstein-Uhlenbeck process from being considered a reliable model for
relative dispersion.
\end{abstract}

\section{Introduction}
Relative dispersion is a process that depends on the combination of
the Eulerian and Lagrangian properties of turbulence. If particle
separation falls in the inertial subrange, the Eulerian spatial structure
affects the dispersion, which can be regarded as a Lagrangian these
property \citep{monin_yaglom-1975}. The combination of properties
requires that both descriptions be considered \citep[see
\eg][]{boffetta_etal-pre-1999}.

Lagrangian Stochastic Modelling (LSM) is one turbulence representation
that naturally combines Eulerian spatial structure and Lagrangian
temporal correlation. In fact, as formulated by \citet{thomson-jfm-1990}
using the Well Mixed Condition (WMC), Lagrangian and Eulerian statistics
are accounted for through the second order Lagrangian structure function
and the probability density function (pdf) of Eulerian velocity. Several
studies prove that this approach leads to the qualitative reproduction
of the main properties, as expected from the Richardson theory
\citep[see][among
others]{thomson-jfm-1990,reynolds-jam-1999,sawford-arfm-2001}.
Furthermore, recent experimental studies seem to confirm the validity of
the Markovianity assumption for the velocity
\citep{laporta_etal-nature-2001,renner_etal-jfm-2001,mordant_etal-prl-2001}.

However, the intrinsic non-uniqueness of the WMC formulation \citep[see,
\eg][]{sawford-blm-1999} and the indeterminacy of the Kolmogorov
constants \citep[see, \eg][for
reviews]{sreenivasan-pf-1995,anfossi_etal-blm-2000} do not allow for a
completely reliable representation of the process. In particular, the
value of the Richardson constant predicted by previous studies is not
uniquely determined \citep[see, among
others,][]{thomson-jfm-1990,kurbanmuradov-mcma-1997,borgas_etal-jfm-1994,reynolds-jam-1999}.
Whether this indetermination is a result of the different formulation of
models, or of the different values of the parameters adopted is still
unclear,
and no systematic studies have been performed so far.

It is worth noting that, even focusing attention only on the dependence
on the model constants produces significant variability. As an example,
\citet{borgas_etal-jfm-1994} present the variation
of the Richardson constant value with the Lagrangian Kolmogorov constant
$C_0$.

The aim of this article is to investigate the general properties of models
based on the WMC with regard to inertial subrange relative dispersion
features. In \Sectref{sect:nondim-WMC} the properties of
the WMC are evidenced through a dimensional analysis, while the limit for
vanishing spatial correlation is studied in \Sectref{sect:OU}.
Subsequently
a model formulation is discussed in \Sectref{sect:model},
and results analysed in \Sectref{sect:results}.

\section{The non-dimensional form of the well mixed condition}\label{sect:nondim-WMC}

Following the logical development of \citet{thomson-jfm-1987},
\citet{thomson-jfm-1990} (hereinafter T90) extended the method for the
selection of single particle Lagrangian Stochastic Models to
models for the evolution of particle pair statistics. In the latter models,
the state of a particle pair is represented by the joint vector of
position and velocity $(\mathbf{x},\mathbf{u})
\equiv(\mathbf{x}^{(1)},\mathbf{x}^{(2)},\mathbf{u}^{(1)},\mathbf{u}^{(2)})$,
where the upper index denotes the particle, whose evolution is given by the
set of Langevin type equations (LE) (with implied summation over repeated indices):
\be
\left \{
\begin{array}{l}
\drm x_i = u_i \drm t \\
\drm u_i = a_i({\bf x},{\bf u},t) \drm t + b_{ij} ({\bf x},t) \drm W_j(t)\,,
\end{array}
\right.
\label{eq:langevin}
\ee
where $i,j=1..6$.
The coefficients $\mathbf{a}$ and $\mathbf{b}$ are determined, as usual,
through the well known Well Mixed Condition
\citep{thomson-jfm-1987} and the consistency with the inertial
subrange scaling, respectively. Further details are not given here, in that
they are well established and widely used in the literature \citep[see,
\eg][for a review]{sawford-arfm-2001}. The only remark we would make is
that, although \citet{thomson-jfm-1987}
himself studied this alternative, the tensor $b_{ij}$ cannot be
dependent on ${\mathbf{u}}$ in order to allow \eqref{eq:langevin} to
describe a physically meaningful process. In fact, as shown, for instance, by \citet{van_kampen-1981},
the It\^{o} and Stratonovich calculus give different results when
$b_{ij}=b_{ij}(\mathbf{u})$. In particular, the WMC would not have a unique definition.
Thus, from now on,
$b_{ij}=\sqrt{C_0 \varepsilon} \delta_{ij}$,
$i,j=1..6$ will be used according to the usual scaling of Lagrangian structure
function \citep{thomson-jfm-1987}, where $\varepsilon$ is the mean dissipation rate of turbulent
kinetic energy.

It should be remembered here that the WMC is satisfied by constraining the
Fokker-Planck equation associated to \eqref{eq:langevin} \citep[see,
\eg][]{gardiner-1990} to be consistent with the Eulerian probability
density function of the flow. In the case of particle pairs the
considered pdf is the one-time, two-point joint pdf of
$\mathbf{x}^{(i)}$ and $\mathbf{u}^{(i)}$, $i=1,2$, accounting for the
spatial structure of the turbulent flow considered. The open question
about the non-uniqueness of the solution in more than one dimension
\citep[see, \eg][]{sawford-blm-1999} is not addressed here. However,
the following analysis will show that the problem studied is
independent of the particular solution selected.

In order to highlight the effect of turbulence features on the model
formulation, characteristic scales for particle pair motion must be
identified. Because the process of relative dispersion has to deal with
both Eulerian and Lagrangian properties \citep[see,
\eg][p. 540]{monin_yaglom-1975}, such scales can be defined
by considering the second order Eulerian and Lagrangian structure
functions, \ie
\be
\mean{\Delta v^2} \sim \subs{C}{K} (\varepsilon \Delta r)^{2/3}
\label{eq:Esf}
\ee
for Eulerian velocity $v$ for a separation $\Delta
r=||\Delta\mathbf{r}||$, according to the standard \citet{kolmogorov-1941} theory
(hereinafter K41), and
\be
\mean{\Delta u^2} \sim C_0 (\varepsilon t)
\label{eq:Lsf}
\ee
for Lagrangian velocity $u$ \citep[see,
\eg][]{monin_yaglom-1975}, where $\Delta
v=||\mathbf{v}(\mathbf{r}+\Delta\mathbf{r})-\mathbf{v}(\mathbf{r})||$ and
$\Delta u=||\mathbf{u}(t+\drm t)-\mathbf{u}(t)||$. A length scale
$\lambda$ can be defined in the Eulerian frame,
so that in the inertial subrange (namely, for $\eta \ll r \ll \lambda$
where $\eta$ is the Kolmogorov microscale) the
structure function for each component may be written as
\be
\mean{\Delta v_i^2}=2 \sigma^2 (\frac{\Delta r}{\lambda})^{2/3}
\label{eq:lambda}
\ee
where $\sigma=\sqrt{||\mathbf{v}||^2/3}$,
which together with \eqref{eq:Esf} provides a definition for $\lambda$.

A Lagrangian time scale $\tau$ can be defined in a similar way using
\eqref{eq:Lsf} and the Lagrangian version of \eqref{eq:lambda}.
Thus, for $\tau_\eta \ll t \ll \tau$, one has
\be
\mean {\Delta u_i^2} = 2 \sigma^2\frac{t}{\tau}
\label{eq:tau}
\ee
from which one can retrieve the known relationship
\be
\varepsilon=\frac{2 \sigma^2}{C_0\tau}
\ee
suggested by \citet{tennekes-1982}.
It should be observed that scales for the inertial subrange,
at variance with their integral version, can be defined independently of
non-homogeneity or unsteadiness, provided that the scales of
such variations are sufficiently large to allow an inertial subrange
to be identified.
As far as the velocity is concerned, $\sigma$
can be recognised as the appropriate
scale of turbulent fluctuations in both descriptions.

The quantities $\sigma$, $\lambda$ and $\tau$ can then be used respectively to make
velocity $u_i$, position $x_i$ and time $t$ non-dimensional.
They also form a non-dimensional parameter
\be
\beta=\frac{\sigma\tau}{\lambda}=\frac{\subs{C}{K}^{3/2}}{\sqrt{2}C_0},
\label{eq:beta}
\ee
the last equality being based on the combination of Eqs.
(\ref{eq:Esf}) and (\ref{eq:Lsf}) with (\ref{eq:lambda}) and (\ref{eq:tau}).
The parameter $\beta$ can be recognised as a version of the well
known Lagrangian-to-Eulerian scale ratio. The approach adopted here
evidences its connection to fundamental constants of the K41
theory.

In non-dimensional form, \eqref{eq:langevin} reads
\be
\left \{
\begin{array}{l}
\drm x_i = \beta u_i \drm t \\
\drm u_i = a_i \drm t + \sqrt{2}\drm W_i(t).
\end{array}
\right.
\label{eq:non_dim_langevin}
\ee
where, with a change of notation with respect to \eqref{eq:langevin},
all the quantities involved are without physical
dimensions.

The associated Fokker-Planck equation is
\be
\frac{\partial p_L}{\partial t}+\beta u_i \frac{\partial p_L}{\partial
x_i} +
\frac{\partial a_i p_L}{\partial u_i} =
\frac{\partial^2 p_L}{\partial u_i \partial u_i}
\label{eq:non_dim_fp}
\ee
where $p_L$ is the pdf of the Lagrangian process
described by \eqref{eq:non_dim_langevin} for some initial conditions.
Using the WMC, $\mathbf{a}$ can be written as
\be
a_i=\frac{\partial \ln p_E}{\partial u_i} + \phi_i
\label{eq:a}
\ee
where
\be
\frac{\partial \phi_i p_E}{\partial u_i}=-\frac{\partial p_E}{\partial
t}
-\beta u_i \frac{\partial p_E}{\partial x_i}
\label{eq:phi}
\ee
and $p_E$ is the Eulerian one-time, two-point joint pdf of
$\mathbf{x}$ and $\mathbf{u}$.

An advantage of this choice of scales emerges clearly in
\eqref{eq:non_dim_fp}. It shows that, given a Eulerian pdf, once the
non-uniqueness problem is solved by selecting a suitable solution to
\eqref{eq:a}, or applying a further physical constraint to
\eqref{eq:phi} \citep{sawford-blm-1999}, any solution of
\eqref{eq:non_dim_fp} will depend on one parameter only, namely on the
Lagrangian-to-Eulerian scale ratio. It can also be observed that this
dependence is completely accounted for by the non-homogeneity term,
which is an intrinsic property of the particle pair dispersion process
in spatially structured velocity fields.

In looking for the universal properties of pair-dispersion in the inertial
subrange, it is useful to rewrite the Richardson $t^3$ law in
non-dimensional form, \ie $\Delta x^2 = g^* \beta^2 t^3$ where
$g^*=2g/C_0$. In this form, the numerical value of the ``normalised''
Richardson constant $g^*$ depends on $\beta$ only. This
dependence is investigated in the following Sections to highlight
the intrinsic properties of the LSM.

\section{The spatial decorrelation limit}\label{sect:OU}

In the limit $\beta\to\infty$, corresponding to a vanishing Eulerian
correlation scale, the non-dimensionalisation defined in the previous
section fails to apply.  However, in this limit, the WMC solution can be
proven to reduce to an homogeneous process (see Appendix). In
particular, selecting a Gaussian pdf will give the Ornstein-Uhlenbeck
(OU) process.  It is worth noting that the OU process has sometimes been
used to describe Lagrangian velocity in turbulent flows, for instance by
\citet{gifford-ae-1982}, who pioneered the stochastic approach to
atmospheric dispersion. The \citet{novikov-1963} model and the NGLS
model \citep[p. 124]{thomson-jfm-1990} are simple applications of this
concept.

Adopting the choices made in the previous Section, but using the spatial
scale defined by $\tau\sigma$ rather than the vanishing $\lambda$ as a
length scale, the OU process equivalent to \eqref{eq:non_dim_langevin}
is described by the non-dimensional set of linear LE
\be
\left \{
\begin{array}{l}
\drm x_i = u_i \drm t\\
\drm u_i = -u_i \drm t + \sqrt{2} \drm W_i
\end{array}
\right.
\ee
where $i=1..6$.
The equations for the relative quantities ($\Delta u_i, \Delta x_i$) can be
obtained from the difference between quantities relative to the first
($i=1,2,3$) and second ($i=4,5,6$) particles. The resulting set of
equations reads
\be
\left \{
\begin{array}{l}
\drm \Delta x_i = \Delta u_i \drm t\\
\drm \Delta u_i = -\Delta u_i \drm t + 2 \drm W_i
\end{array}
\right.
\label{eq:OU-relative}
\ee
where $i=1..3$.

Equation (\ref{eq:OU-relative}) can be solved analytically for
correlation functions and variances \citep[see \eg][]{gardiner-1990}.
Some basic results are summarised below \citep[see
also][]{gifford-ae-1982}.

The second order moment of velocity difference turns out to be an
exponential function dependent on the time interval only
\be
\mean{(\Delta u_i - \Delta u_{0i})^2}=\mean{\Delta
u_{0i}^2}\exp{(-t)}\,.
\label{eq:OU-du}
\ee
By integrating \eqref{eq:OU-du}, the displacement variance for a single
component is
\be
\mean{(\Delta x_i - \Delta x_{0i})^2}=(\mean{\Delta u_{0i}^2}-2)(1-\exp{(-t)})
+4t-4(1-\exp{(-t)})\,.
\label{eq:OU-dx}
\ee
For short times (but expanding \eqref{eq:OU-dx} to the third power of
$t$), it turns out that
\be
\mean{(\Delta x_i - \Delta x_{0i})^2}\simeq\mean{\Delta
u_{0i}^2}t^2+\left(\frac 4 3 -\mean{\Delta u_{0i}^2}\right)t^3\,.
\label{eq:OU-x-asympt}
\ee
From \eqref{eq:OU-x-asympt} it can be observed that, when initial
relative velocity $\Delta u_{0i}$ is distributed in equilibrium with
Eulerian turbulence (\ie $\mean{\Delta u_{0i}^2}=2$), a $t^2$ regime
takes place with a negative $t^3$ correction
\citep{hunt-arfm-1985}. On the other hand, if $\mean{\Delta u_{0i}^2}=0$
the ballistic regime displays a $t^3$ growth with a coefficient 4, \ie
$2C_0$ for the dimensional version
\citep{novikov-1963,monin_yaglom-1975,borgas_etal-jfm-1991}.

\section{Model formulation and numerical simulations}\label{sect:model}

In order to proceed with the analysis of the dependence of model
features on parameter $\beta$, we select as a possible solution to
\eqref{eq:a}, the expression given by T90 (his eq. 18) for Gaussian pdf.
The spatial structure is accounted for using the \citet{durbin-jfm-1980}
formula for longitudinal velocity correlation, which is compatible with
the 2/3 scaling law in the inertial subrange. Although this form is
known not to satisfy completely the inertial subrange requirements (it
prescribes a Gaussian distribution for Eulerian velocity differences, while
inertial subrange requires a non-zero skewness), it has been successfully
used in basic studies \citep{borgas_etal-jfm-1994} and applications
\citep{reynolds-jam-1999}, and provides a useful test case for studying the
results shown above.

The stochastic model is formulated for the variable
$(\mathbf{x},\mathbf{u})$ rather than for the variable
$(\Delta\mathbf{x}/\sqrt{2}$,$\Delta\mathbf{u}/\sqrt{2})$ as in
Thomson's original formulation. In the present case,
assuming homogeneous and isotropic turbulence, the covariance matrix
$\mathcal{V}(\mathbf{x})$ of the Eulerian pdf is expressed by
\be
\mathcal{V}=\left(
\begin{array}{cc}
\mathcal{I}&\mathcal{R}^{(1,2)}(\mathbf{x})\\
\mathcal{R}^{(2,1)}(\mathbf{x})&\mathcal{I}\\
\end{array}
\right)
\ee
where $\mathcal{I}$ is the identity matrix and
\be
\mathcal{R}^{(p_1,p_2)}_{ij}(\mathbf{x})=\mean{u^{(p_1)}_i u^{(p_2)}_j}
\ee
where $p_1,p_2=1,2$ ($p_1\neq p_2$) are the particles indices.
The quantity $\mean{u^{(p_1)}_i u^{(p_2)}_j}\equiv
\mean{u_i(\mathbf{x}^{(p_1)})u_j(\mathbf{x}^{(p_2)})}$ is the two-point
covariance matrix, which is expressed in terms of longitudinal and
transverse functions $F$ and $G$ \citep[see, \eg][]{batchelor-1970} as
\be
\mathcal{R}_{ij}=F(\Delta_x)\Delta x_i\Delta
x_j+G(\Delta_x)\delta_{ij}
\ee
where $\Delta x=||\mathbf{x}^{(1)}-\mathbf{x}^{(2)}||$,
\be
F=-\frac{1}{2\Delta_x}\frac{\partial f}{\partial\Delta_x}
\ee
and
\be
G=f+\frac{\Delta x}{2}\frac{\partial
f}{\partial\Delta x}\,.
\ee
It goes without saying that $\mathcal{R}^{(p_1,p_2)}_{ij}=
\mathcal{R}^{(p_2,p_1)}_{ij}= \mathcal{R}^{(p_1,p_2)}_{ji}$.
As in \citet{durbin-jfm-1980}, $F$ and $G$ are computed from the parallel
velocity correlation
\be
f(\Delta x)=1-\left(\frac{\Delta x^2}{\Delta x^2+1}\right)^{1/3}\,.
\ee
which is K41 compliant for $\Delta x\ll 1$.

Using the above formulation, \eqrefs{eq:non_dim_langevin} were solved
numerically for a number of trajectories large enough to provide
reliable statistics for the relevant quantities.
Particular attention was paid to the
time-step--independence of the solution (details are not reported here).
It was found that the time step strongly depends on $\beta$ because large
values of the parameter increase non-homogeneity, which requires greater
accuracy. Despite the widespread use of
variable--time-step algorithms \citep[see,
\eg][]{thomson-jfm-1990,schwere_etal-cg-2002} based, in particular, on
spatial derivatives, here a fixed time step short enough for time-step
independence of the solution was used throughout the computation.

Simulations were performed for two different initial conditions for
velocity difference: i) the \textit{distributed} case, where velocity
differences are given according to the second-order Eulerian structure
function and ii) the \textit{delta} case ($\mean{\Delta u_i^2}_0=0$), where
both particles of a pair are released with the same velocity, which
is normally distributed with variance $1$. The two cases correspond to
the limiting cases considered in \Sectref{sect:OU}. The former describes
``real'' fluid particles, \ie particles distributed like fluid at all
times, while the latter represents, from the point of view of relative
dispersion, marked particles leaving a ``forced'' source, where they were
completely correlated (as for a jet). The initial condition
for the spatial variable was $\Delta x_0=10^{-5}\beta$ for all simulations.
It can be noted that this corresponds to different positions in the
inertial subrange for different simulations ($\Delta x_0$ differs from
case to case). However, the dimensional $\lambda \Delta x_0$ is chosen
small enough to provide at least three decades of inertial subrange.

The $\beta$ parameter was varied in the range $[10^{-2}:10^2]$, well
beyond physically meaningful values. In fact, values reported in
the literature range from O($10^{-1}$) to O($10^1$)
\citep{hinze-1959,hanna-jam-1981,sato_etal-jfm-1987,koeltzsch-ae-1999}
with $\beta=O(1)$ taken as a reference \citep{corrsin-jas-1963}. This
choice was made in order to infer asymptotic properties of the model.
Note that, from a numerical point of view,
different values of $\beta$ were obtained by varying the length scale
$\lambda$, keeping $\sigma$, $\tau$ and $C_0$ fixed. In other words,
with reference to \eqref{eq:beta}, the variation of $\beta$ was
obtained by varying $\subs{C}{K}$.

\section{Results and discussion}\label{sect:results}

\newlength{\offset}
\setlength{\offset}{0.5cm}
\newlength{\third}
\setlength{\third}{0.33333\textheight}
\addtolength{\third}{-\offset}
\begin{figure}
\begin{center}
\includegraphics[height=\third]{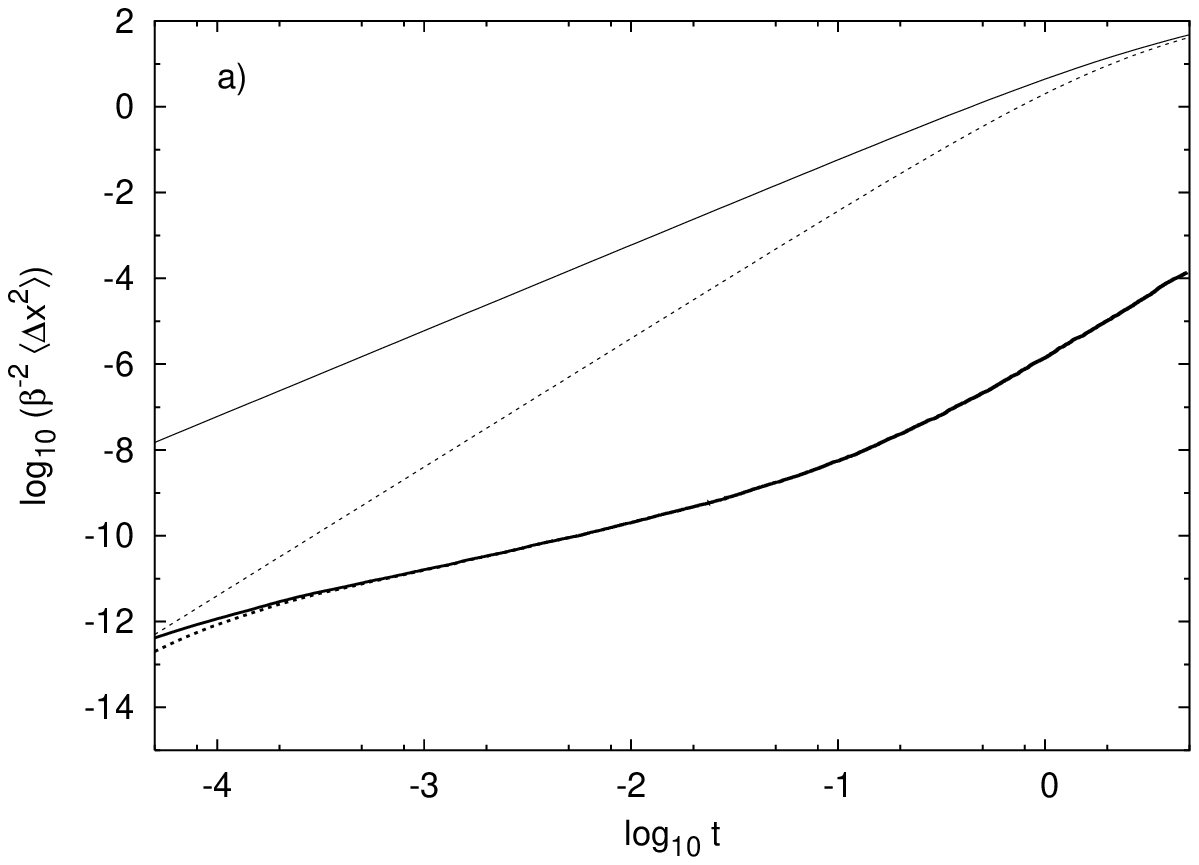}
\includegraphics[height=\third]{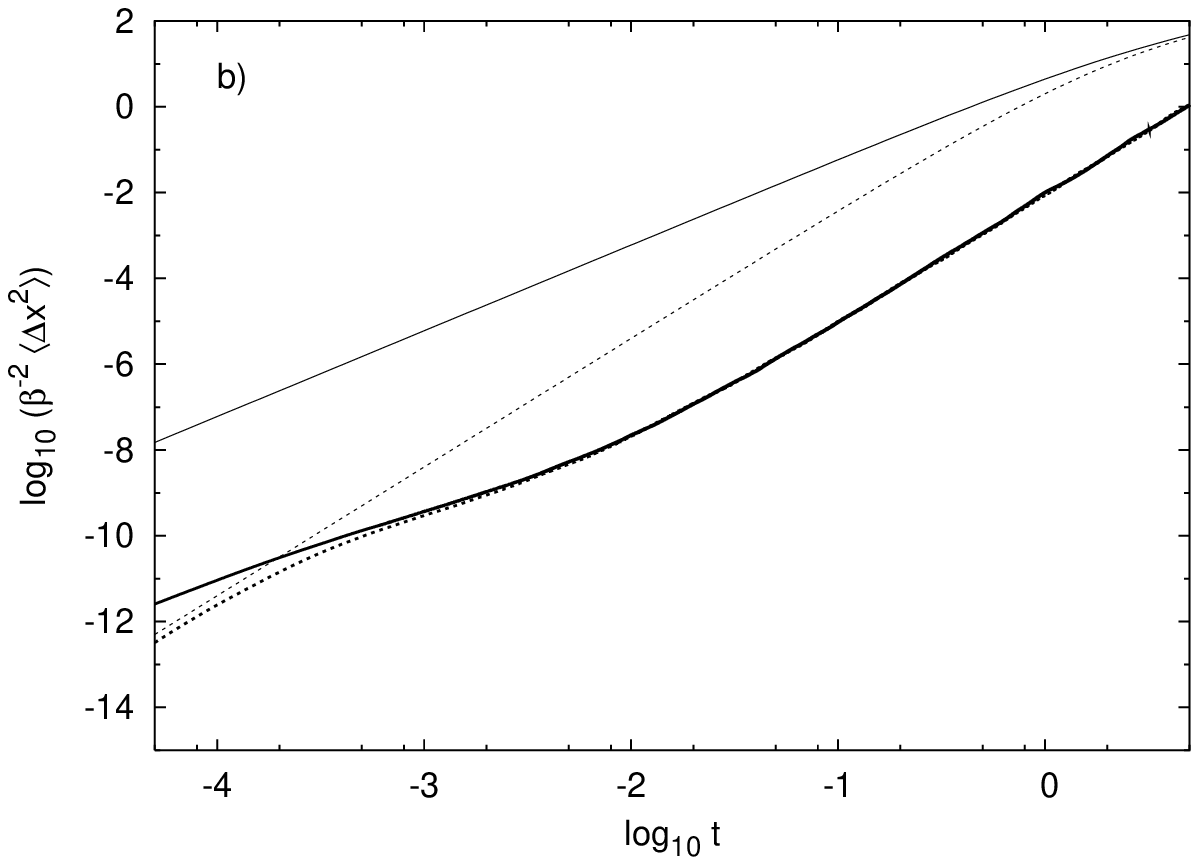}
\includegraphics[height=\third]{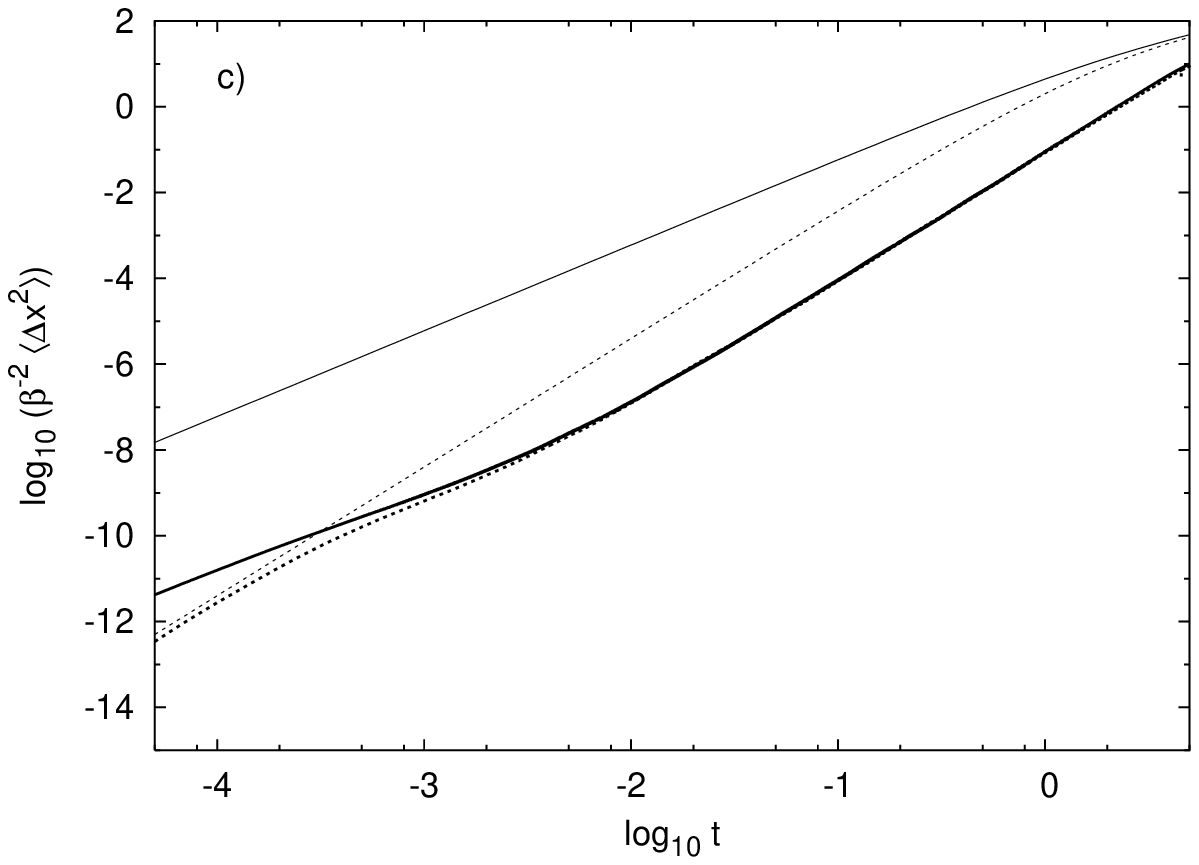}
\caption{Mean square separation normalised with $\beta$ as a function of
time for different values of $\beta$. Thick lines
represent results of present simulations, while thin lines
are the analytical Ornstein-Uhlenbeck solutions (continuous:
\textit{distributed} case; dotted: \textit{delta} case). 
a) $\beta=0.01$, b) $\beta=0.1$, c) $\beta=0.2$.}
\label{fig:1}
\end{center}
\end{figure}
\addtocounter{figure}{-1}
\begin{figure}
\begin{center}
\includegraphics[height=\third]{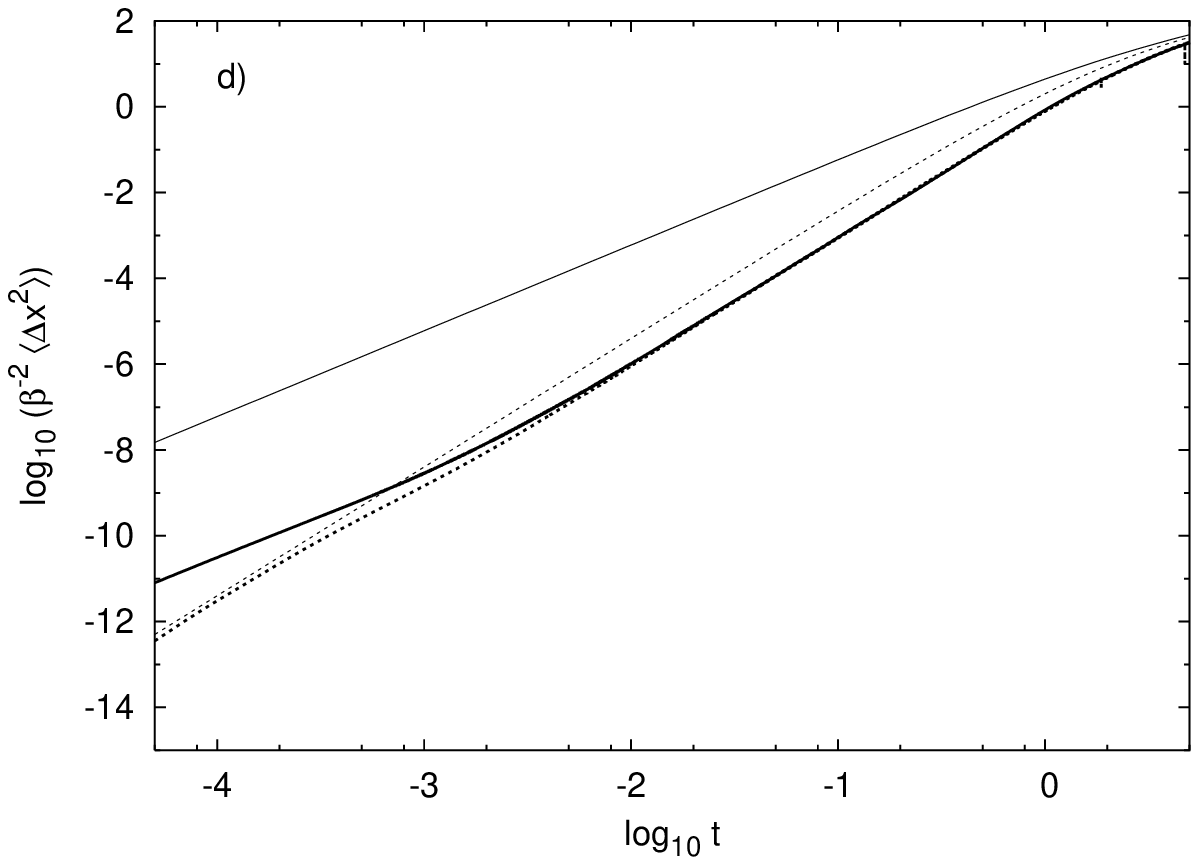}
\includegraphics[height=\third]{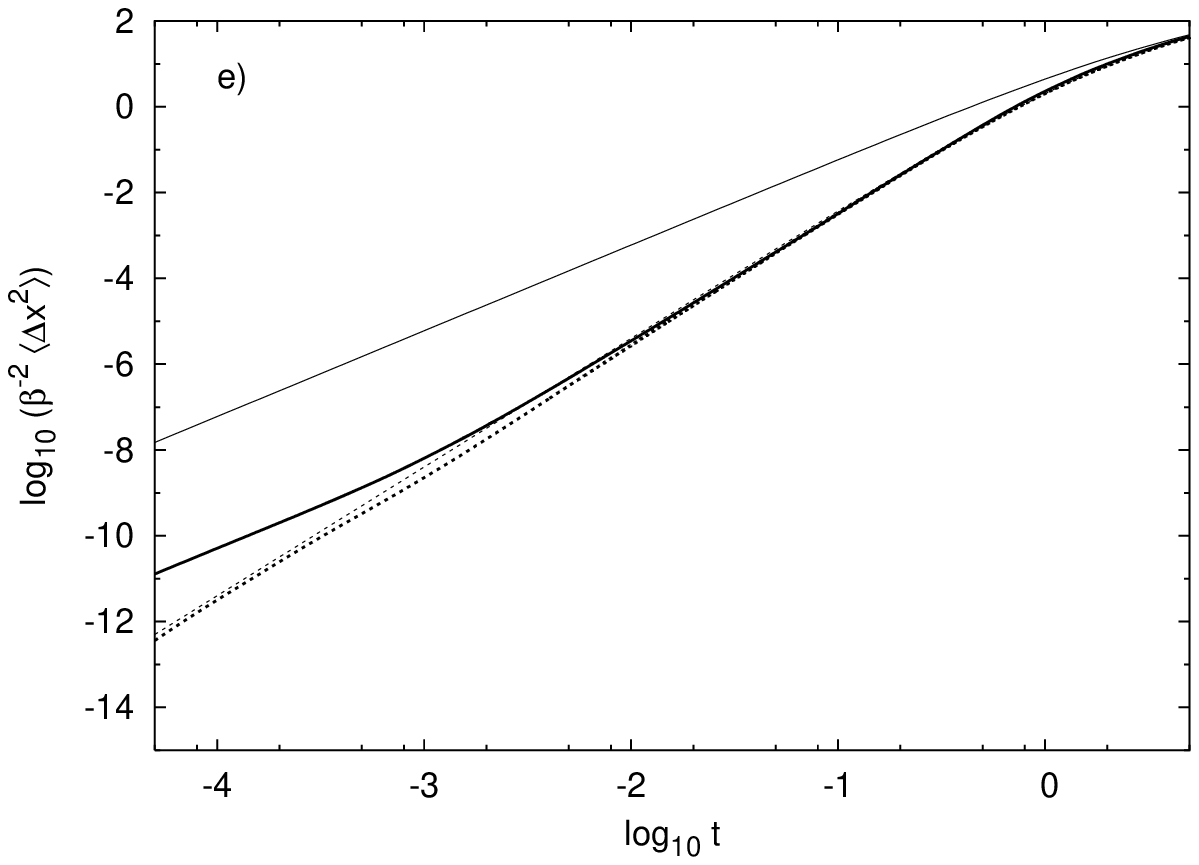}
\includegraphics[height=\third]{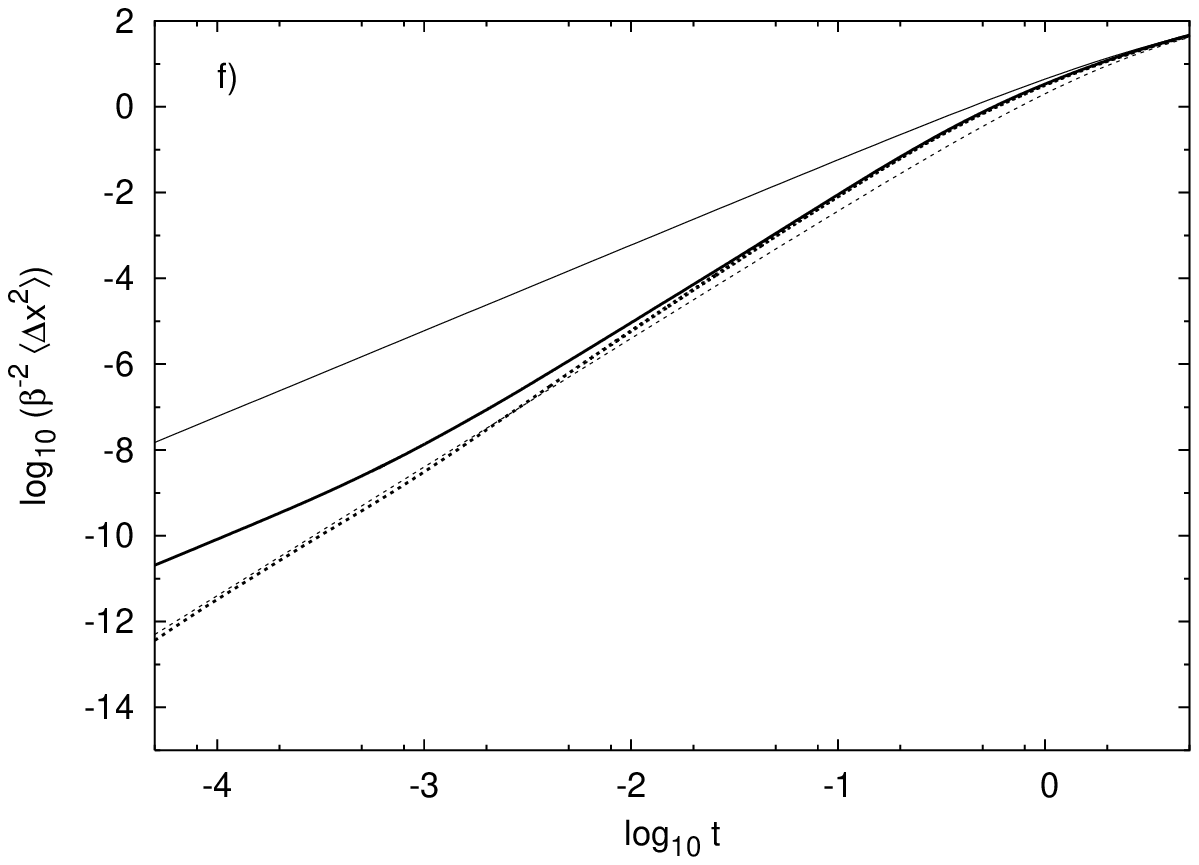}
\caption{(continued) d) $\beta=0.5$, e) $\beta=1$, f) $\beta=2$.}
\end{center}
\end{figure}
\addtocounter{figure}{-1}
\begin{figure}
\begin{center}
\includegraphics[height=\third]{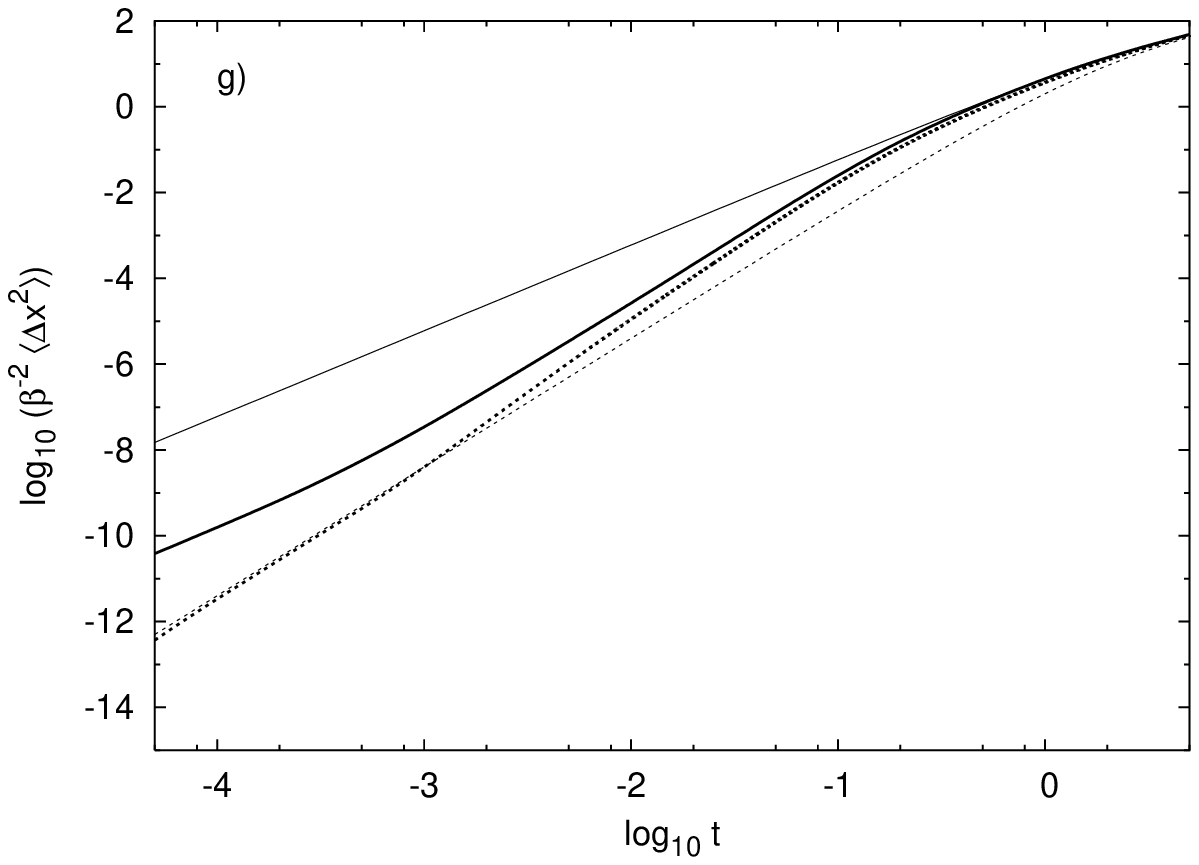}
\includegraphics[height=\third]{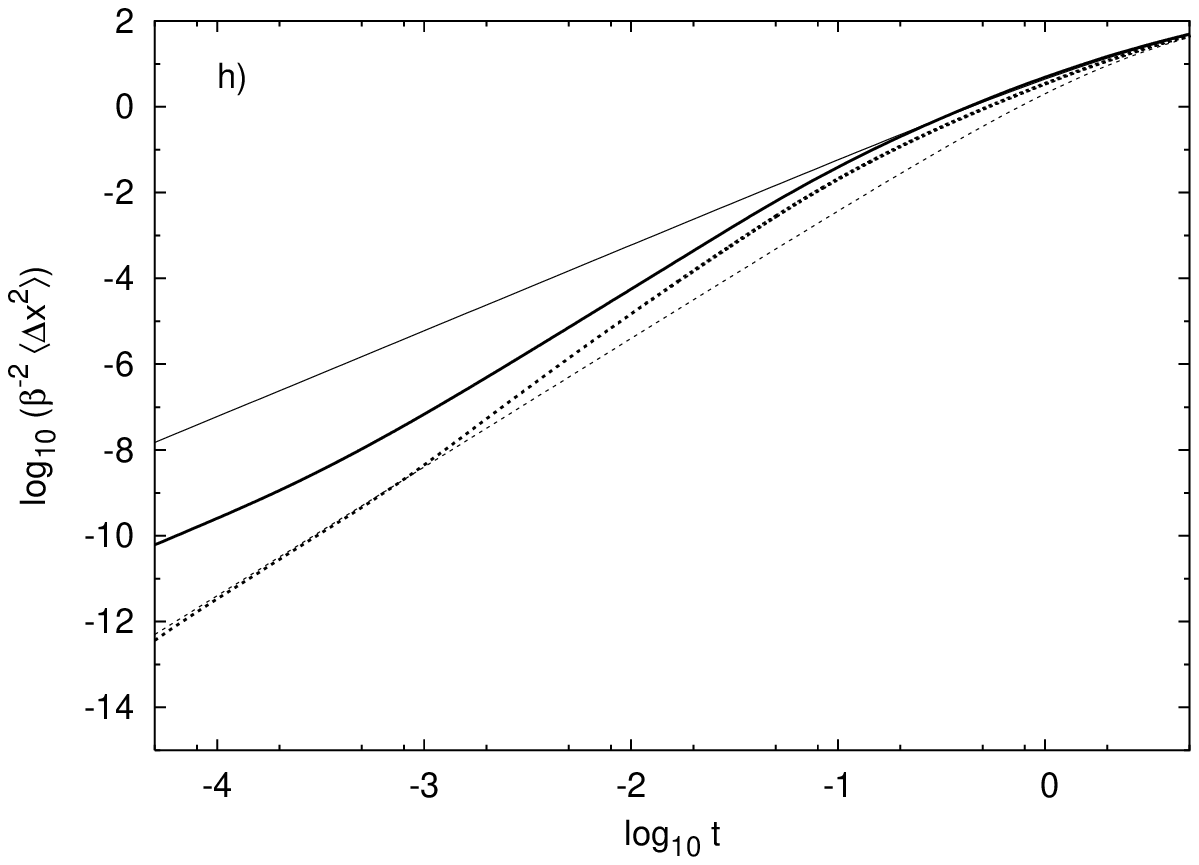}
\includegraphics[height=\third]{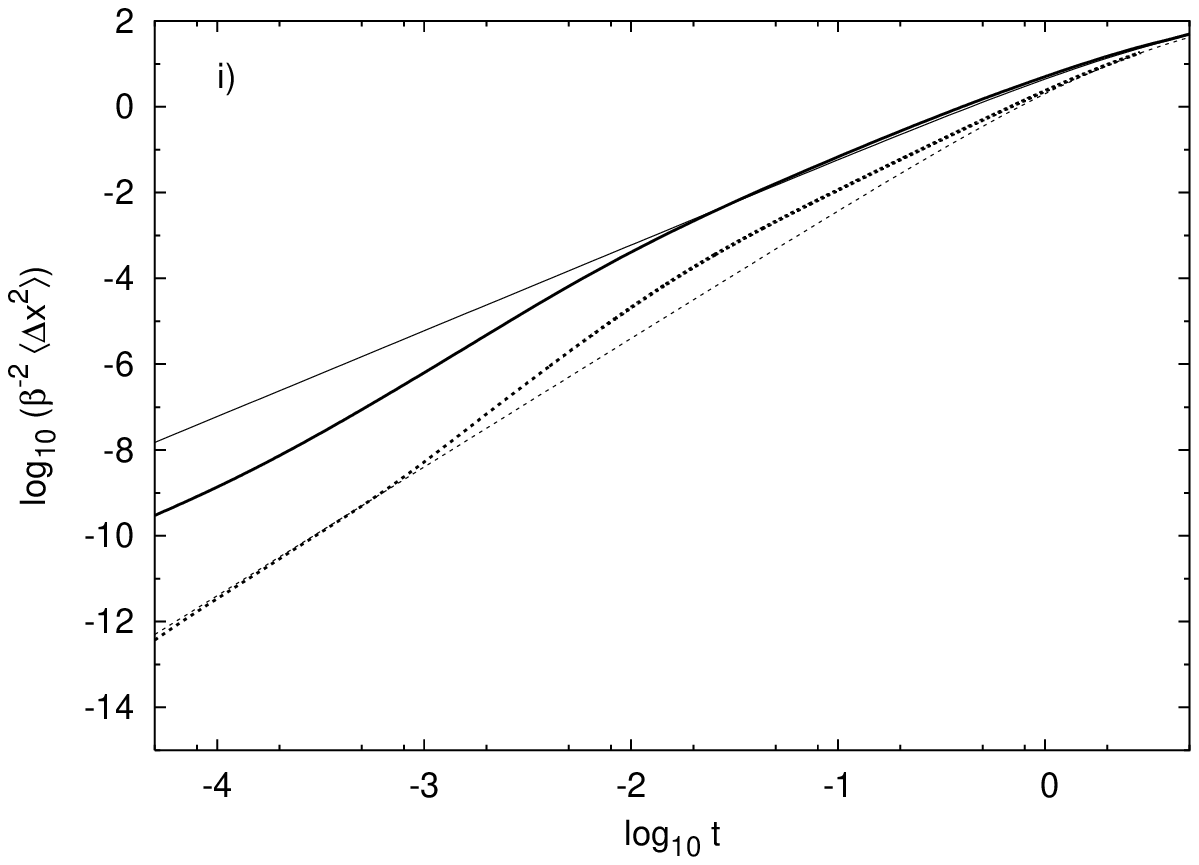}
\caption{(continued) g) $\beta=5$, h) $\beta=10$, i) $\beta=100$.}
\end{center}
\end{figure}

\begin{figure}
\begin{center}
\includegraphics[height=\third]{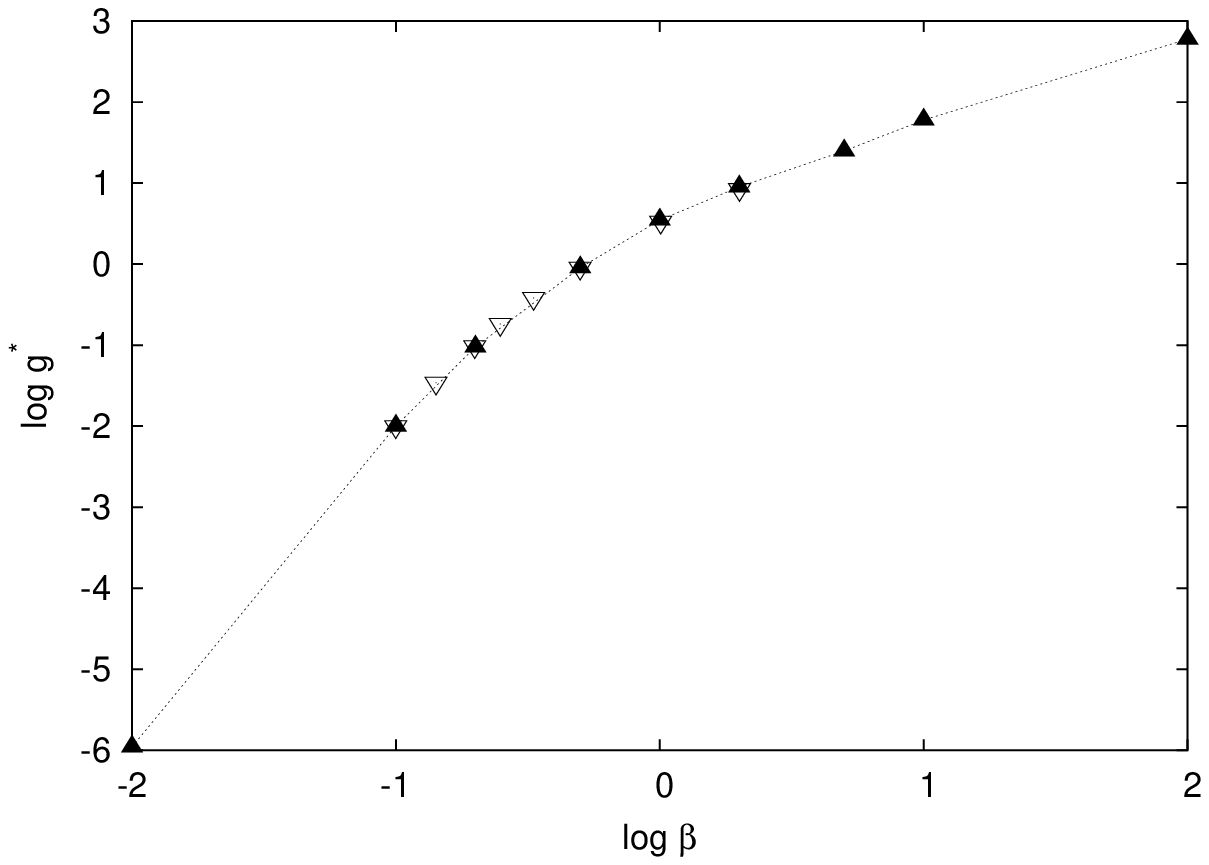}
\caption{Normalized Richardson coefficient \textit{vs.} $\beta$.
Present results are represented by $\blacktriangle$ connected with a
line while
$\triangledown$ are taken from \citet{borgas_etal-jfm-1994}. }
\label{fig:g-star}
\end{center}
\end{figure}

\Figrefs{fig:1} (from a to i) show the results of simulations for the two initial
conditions and for different values of $\beta$. The non-dimensional
quantity $\mean{\Delta x^2}\beta^{-2}$ is plotted against the
non-dimensional time $t$. The OU analytical solutions ($\beta=\infty$) are
reported for reference. The general behaviour qualitatively filfills the
expectations of \citet{taylor-1921} and \citet{richardson-1926}.
It presents an initial ballistic regime which differs for the
two cases: the \textit{distributed} case shows a $t^2$, while the
\textit{delta} case
presents a ``false''\footnote{In the sense that it is
only a correction to the ballistic $t^2$ regime, which depends on the
initial conditions and not on spatial structure.} $t^3$
according to \eqref{eq:OU-x-asympt}. After the ballistic regime there
is a transition to an inertial range $t^3$ regime, which then becomes well
established until a pure diffusive regime takes place.

This generically correct behaviour merits further consideration.
A ``true'' $t^3$ is observed, which depends on the
spatial flow structure and influences dispersion properties.
In particular, increasing $\beta$ causes an increase in the
normalised Richardson coefficient $g^*$ (\figref{fig:g-star}). It is
worth noting that the ``false'' $t^3$ regime, according to
the findings reported in \sectref{sect:OU}, is not dependent on the
structure, and therefore does not vary with $\beta$. In fact, as pointed
out by \citet{sawford-arfm-2001}, there should be a range where
$\mean{\Delta x^2}\beta^{-2}=4t^3$
for $t\ll t_0$, $t_0$ being the time at which memory of the initial
conditions is lost \citep[see also][]{borgas_etal-jfm-1991}. It is clear
now that
this regime does not originate from any spatial structure and is
intrinsic to the solution with the \textit{delta} initial condition, as
explained by \eqref{eq:OU-x-asympt}.

According to \citet[p. 541]{monin_yaglom-1975}, the ``true'' $t^3$
regime should be independent of the initial conditions. Thus, the
starting point of this regime can be selected at the point where the
solutions for the two cases coincide, as clearly occurs in
\figrefs{fig:1}d to f.  Therefore, the temporal extension of the $t^3$
regime is probably shorter than the one that could be estimated using
intersections with the idealised ballistic regime, on the one hand, and
with the diffusive regime, on the other.  Note, however, that the
extension of the inertial regime remains a decreasing function of
$\beta$, which asymptotically converges to zero.

Another point of interest evident in \figrefs{fig:1} and 2
is that the present results do not agree with the
theoretical findings of \citet{borgas_etal-jfm-1991} (hereinafter BS91),
although they compare well with the numerical results of
\citet{borgas_etal-jfm-1994} (hereinafter BS94).
In fact, BS94 (their last figure) showed the results obtained by varying $C_0$
in their models. However, as shown in \sectref{sect:nondim-WMC}, $\beta$
is the only parameter on which the model depends. Because of the
constancy of $\varepsilon$ in BS94, the varation of $C_0$
corresponds to a variation of $\tau$, and hence $\beta$. The results of
BS94 for the implementation of the T90 model are reported in
\figref{fig:g-star}, and show a complete agreement with the present results.
The values indicated do not satisfy the kinematic constraint $g=2C_0-\gamma$,
where $\gamma$ is a positive quantity, proposed by BS91, based on a
double asymptotic expansion.  It should be observed, however, that
$\gamma$ is derived from kinematic features and depends on integrals of
correlation functions. For vanishing correlation, one obtains
$\gamma\to0$,
suggesting that in BS91 the ballistic part of the OU process is an upper
limit for dispersion in the Richardson regime.

Nevertheless, this discrepancy can be explained as follows. From
\eqref{eq:OU-x-asympt} it is clear that, at any time
$t<t_{\mbox{\scriptsize{diff}}}$, where $t_{\mbox{\scriptsize{diff}}}$ is
the time when pure diffusion takes place, the displacement variance for
the OU process in the \textit{distributed} case is always larger than
the displacement variance for the OU process in the \textit{delta} case.
The two cases represent the limit for any process based on the WMC for
$\beta\to\infty$, as shown in \sectref{sect:OU}. In particular, focusing
attention on the time range between the ballistic and diffusive regimes,
it can be
observed that $\lim_{\beta\to\infty} \mbox{T90}(distributed) =
\mbox{OU}(distributed)$ and, furthermore, $\mbox{T90}(distributed) <
\mbox{OU}(distributed)$ for any finite $\beta$. It can be concluded
that a $\beta^\prime$ must exist for which $\mbox{T90}(distributed)=
\mbox{OU}(delta)$, and 
$\mbox{T90}(distributed) > \mbox{OU}(delta)$ for $\beta>\beta^\prime$,
in disagreement with BS91. Recalling that the OU process is the limit for
any WMC process with Gaussian $p_E$, this result can be considered to be applicable to more
general kinematic properties, which should therefore depend on the
ratio between Eulerian and Lagrangian scales.
Thus, the limitation to $g^*$ in the BS91 derivation possibly derives from
an implicit assumption concerning the spatial structure and/or the value of
$\beta$, which defines a range of applicability of the result.

Proceeding further with the analysis, it can also be said that, because
of the existence of a time $t_0$ after which the solution is not dependent
on the initial
conditions, it might be expected that $\mbox{T90}(distributed) =
\mbox{T90}(delta)$ for $t>t_0$.
However, as the Lagrangian time increases with respect to
$\sigma^{-1}\lambda$, an increasing number of particle pairs
reaches the end of the inertial range ($\Delta x\gg1$)
still remembering their initial
conditions.
This results in a range of $\beta > 1$ in which the \textit{delta}
solution never reaches the \textit{distributed} solution before the onset
of the diffusive regime.
Therefore, it is not possible to define any $g^*$.
Nevertheless, for $\beta>\beta^\prime$ there exists a range of $t$
where $\mbox{T90}(delta)>\mbox{OU}(delta)$, which shows that $\mbox{T90}(delta)$
converges to $\mbox{OU}(delta)$ in a non-monotonic way.

When, for $\beta \lesssim 1$, the expected independence on the initial
conditions is recovered, it can be noted that $t_0$ itself is a function
of $\beta$.  Thus, the duration of the $t^3$ regime depends also (and
mainly) on the starting time of the diffusive regime It is observed that
decreasing $\beta$ increases the time at which the diffusion regime
becomes fully developed.

\section{Conclusions}

The dimensional analysis of the WMC, through the non-dimensionalisation
of the Fokker-Planck equation has shown that only one parameter plays a
role in the determination of two particle dispersion properties.
This parameter is the Lagrangian-to-Eulerian scale ratio $\beta$, which can
be reliably defined in terms of inertial subrange constants.
The dimensional analysis leads to the definition of a normalised
Richardson constant $g^*$ whose scale is identified with $C_0$, as
suggested by the comparison of Lagrangian and Eulerian properties.
Given a particular model, the numerical value of $g^*$ depends
solely on the value of $\beta$ adopted. This also applies to the
duration of the $t^3$ regime.

Using the T90 formulation, it has been shown that the results of
\citet{novikov-1963} are recovered for $\beta\to\infty$, which means
that in the model the spatial structure is negligible with respect to
the Lagrangian time correlation. This limit corresponds to the OU
process, whose general properties highlight that the observed $t^3$
growth is actually a correction to the ballistic regime $t^2$.
Moreover, because of the absence of any genuine $t^3$ regime, it is not
possible to define any Richardson coefficient. This
means that $2C_0$ cannot be considered in general as the upper limit for
$g$.  Therefore there is no inconsistency in models that produce
$g>2C_0$, as occurs in the present study and in BS94.

\section*{Acknowledgements}
G. Pagnini is supported by the CNR-fellowship n. 126.226.BO.2.

\bibliographystyle{personal}
\bibliography{abbr,all}

\renewcommand{\theequation}{A-\arabic{equation}}
\setcounter{equation}{0}  
\section*{Appendix}  

The stationary structure function of the second order,
\eqref{eq:lambda}, can be generalized to an arbitrary
integer order $n$, in non-dimensional terms as
\be
\mean{\Delta u^n} = \subs{\mean{\Delta u^n}}{e} \Delta r^{hn},
\label{s-s}
\ee
where $\subs{\mean{\cdot}}{e}$ denotes Eulerian equilibrium statistics
and, when $n=2$, $\subs{\mean{\Delta u^2}}{e}=2$.
The inertial subrange and spatial decorrelation limit are recovered for
$h=1/3$ and $h=0$, respectively.

Considering the characteristic function $\hat{p}_E(\Delta w;\Delta r)$
of the stationary Eulerian $pdf$ of velocity differences $p_E(\Delta
u;\Delta r)$ and using \eqref{s-s}, it turns out that
\be
\hat{p}_E(\Delta w;\Delta r)=
\sum_{n=0}^{\infty} (i \Delta r^h \Delta w)^n \subs{\mean{\Delta
u^n}}{e}(n!)^{-1} =
\hat{f}(\Delta r^h \Delta w),
\label{s-s-pdf}
\ee
with $i=\sqrt{-1}$. From \eqref{s-s-pdf} it follows that
\be
p_E(\Delta u; \Delta r)
= \frac{1}{\Delta r^h} f \left(\frac{\Delta u}{\Delta r^h}\right) \,,
\label{A-pdf}
\ee
where the factor $\Delta r^{-h}$ conserves the normalization and, for the
constant values $h=1/3,0$, \eqref{A-pdf} defines the self similar
regimes of the inertial subrange and the spatial decorrelation limit,
respectively.

Using the dimensional quantities $\Delta r^\prime=\lambda \Delta r$ and 
$\Delta u^\prime=\sigma \Delta u$ for the particle separation and the velocity
differences, respectively, for any finite Lagrangian correlation time $\tau$
and particle separation $\Delta r^\prime$, the following identity holds
\be
\lim_{\beta \to \infty} \varphi(\Delta r) \equiv 
\lim_{\lambda \to 0} \varphi(\Delta r^\prime /\lambda)\,,
\ee
where $\varphi$ is a generic continuous bounded function.
Since continuity is required in the transition from the inertial subrange
regime to the equilibrium, the scaling exponent $h$ is assumed to be a
monotonic decreasing function of $\Delta r^\prime/\lambda$. Thus
\be
\lim_{\lambda \to 0}  
\lambda^h = 1 \,.
\label{l=1}
\ee

As observed in \Sectref{sect:nondim-WMC}, the only term affected by
variations of $\beta$ in \eqref{eq:non_dim_fp} is the non-homogeneous
one. Therefore for any finite $\Delta r^\prime$ using
\eqref{A-pdf} and \eqref{l=1}, it turns out that
\begin{eqnarray}
\lim_{\beta \to \infty} \, \beta \, \frac{\partial p_E}{\partial r} \sim 
\lim_{\lambda \to 0} 
&&\left\{
\lambda^h 
\frac{h}{\Delta {r^\prime}^{h+1}} 
f \left( \frac{\Delta u^\prime \sigma^{-1}}{(\Delta r^\prime \lambda^{-1})^h}\right) + 
\right.
\nonumber \\
&&\lambda^{2h} \left. \frac{h}{\Delta {r^\prime}^{2h+1}} \frac{\Delta
u^\prime}{\sigma} 
f^\prime \left(\frac{\Delta u^\prime \sigma^{-1}}{(\Delta r^\prime \lambda^{-1})^h}\right)
\right\}
\rightarrow 0
\end{eqnarray}
which shows that the non-homogeneous term vanishes in this limit.

\end{document}